\documentclass[preprint2]{aastex62}
\submitjournal{AJ}

\shorttitle{Quadruply lensed quasar taxonomy}
\shortauthors{Schechter}

\begin{document}

\title{A TAXONOMY FOR THE CONFIGURATIONS OF QUADRUPLY LENSED QUASARS}

\correspondingauthor{Paul L.\ Schechter}
\email{schech@mit.edu}

\author[0000-0002-5665-4172]{Paul L.\ Schechter}
\affil{MIT Kavli Institute 37-635,\\
  77 Massachusetts Avenue, Cambridge, MA, 02138-4307, USA}\



\begin{abstract}
  A simple, novice-friendly scheme for classifying the image
  configurations of quadruply lensed quasars is proposed.  With only
  six classes, it is intentionally coarse-grained.  It focuses on the
  kitelikeness and circularity of these configurations, or the absence
  thereof.  Other features are deliberately ignored,
  their importance to professional astronomers notwithstanding. 
  Readers are invited to test drive the scheme on a sample
  of 12 quadruply lensed quasar systems.  The theoretical
  underpinnings of the scheme are described in a technical appendix.
\end{abstract}



\section{Introduction}\label{sec:intro}

Astronomers who study quadruply lensed quasars (henceforth
``experts''), can, by visual inspection, reliably discriminate between
random quartets of stars and the four images of a single lensed
source.  The four celestial positions of the images lie in an
8-dimensional space.  Random quartets fill that space far more
uniformly than do lensed quasar images, which lie in a limited region
of an approximately 7-dimensional subspace defined by the
``configuration invariant'' discovered by \citet{KassiolaKovner}.

Four of the seven dimensions are uninteresting, representing
translations, rotations and scalings of otherwise identical
configurations.  The experts ignore these four, leaving three
unitless quantities (angles and ratios of distances) that
describe the salient features of the quadruple configurations,

The premise of this paper is that the three dimensional description of
quadruple lensed image configurations is sufficiently simple that
even novices can quickly master it.

While citizen scientists will doubtless want to develop an
understanding of the theory behind the phenomenon, it can be distilled to
basic ideas: a) that quadruple image configurations often resemble
kites and b) the configurations are sometimes non-circular and other
times nearly circular.

\begin{figure*}[t]\label{fig:thescheme}
  \centering
   \includegraphics[scale=0.85, angle=90]{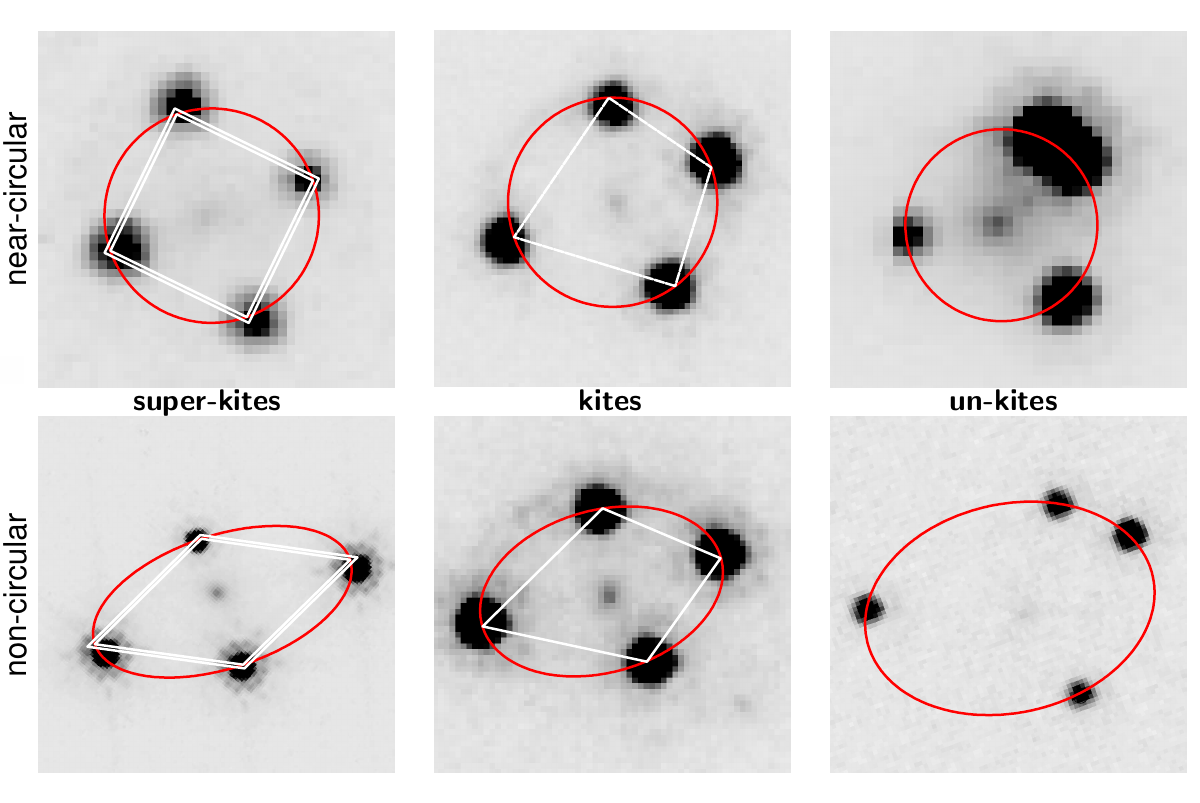}
    \caption{Classification scheme for the configurations of quadruply
      lensed quasars.  The white lines show the results of fitting
      theoretical models of perfect kites and super-kites to the image
      positions.  The models for the top row were constrained to be
      circular.  The red circles (top row) and ellipses (bottom row)
      show the loci of image positions allowed by each model.  The
      models are described in Appendix \ref{app:expert}.  The systems
      in the top row are J2344$-$30, J0259$-$16 and HS 0810+25.  On
      the bottom row are J1134$-$21, J1606$-$23 and J2033$-$47.}
\end{figure*}

In distancing our scheme from theory, we have followed the example
of the stellar spectral classification system.  Not only is an
understanding of effective temperature and surface gravity unneeded,
one might argue that the scheme is useful to professional astronomers
precisely because it steps back from theory.

The classification scheme is described in Section \ref{sec:thescheme}.
In Section \ref{sec:ignore} we discuss observational features
of quadruply lensed systems that we choose to ignore.
In Section \ref{sec:ellipses} we explain why non-circularity
is more difficult to distinguish than kitelikeness.
In Section \ref{sec:testdrive} we invite readers to
classify a sample of a dozen systems. 
Appendix \ref{app:expert}
contains additional material for experts.
In Appendix \ref{app:encircle}
we give the results of theoretical fits
of a singular isothermal elliptical 
potential to each of the sample systems.

\section{The Scheme}\label{sec:thescheme}

We ask three questions about the four images,  
all qualified with the words ``more or less":
\vskip10pt
{\par
{\parindent=25pt
\obeylines
Do the four images trace the outline
of a kite?
~~~~If so, are the four images symmetric
~~~~around {\it both} diagonals?
~~~~~~~~If so, the system is a ``super-kite.''
~~~~~~~~If not, the system is a ``kite.''
~~~~If not, the system is an ``un-kite.''
Do the four images lie on a circle?
~~~~If so, the system is ``near-circular.''
~~~~If not, the system is ``non-circular.''
\par
}}
\vskip10pt
In Figure 1 we show images obtained with the Hubble Space Telescope
(henceforth HST) of representatives of each of our six classes.  Both
conceptually and graphically, ``kites" are at the center of our
scheme.  These are quadruple image configurations that are
approximately symmetric around one of their two diagonals.

Kites vary in the degree of asymmetry around the other diagonal, but
we ignore this and call all such systems kites.  The exception is that
we identify as ``super-kites" systems that are very nearly symmetric
around both diagonals.  Systems that are asymmetric around both
diagonals are classified as ``un-kites."

Each of these three classes is further split by whether or not the
four quasar images lie on a circle, giving ``near-circular'' and
``non-circular'' systems.

Prospective users of this scheme are {\it discouraged} from thinking
too carefully about the distinctions, and are encouraged, instead,
to choose whichever class seems ``more nearly" to describe a
particular quadruply lensed quasar configuration.

Users of the scheme are {\it not} expected to measure positions and
fit theoretical models to the positions of the four images.  We
nonetheless show ellipses in Figure \ref{fig:thescheme} that result
from fitting our preferred theoretical model, the singular isothermal
elliptical potential (henceforth SIEP; described further in Appendix
\ref{app:expert}) to show that non-circularity can, for this model, be
precisely quantified.

\section{Features to ignore}\label{sec:ignore}

Our three dimensional scheme ignores observable features of
gravitationally lensed quasars that are nonetheless of considerable
astronomical interest.  The physics that governs these does not affect
image positions and can be quite complicated.  Expert users will want
to take note of these, but for the purposes of our scheme, they are
distractions.

\subsection{Ignore the fluxes}\label{subsec:fluxes}
We deliberately ignore the relative fluxes of the quasar images in our
taxonomy as they can undergo microlensing fluctuations of two
magnitudes or more \citep{Weisenbach_2021}.  But they can and should
be taken into account in deciding {\it whether} a quartet of images is
a lensed quasar.  Close pairs of lensed quasar images tend to be brighter
than those that are more widely separated, more so in near-circular
systems \citep{FalorSchechter}, as in the upper right panel of
Figure \ref{fig:thescheme}.
\begin{figure}[h!]
\plotone{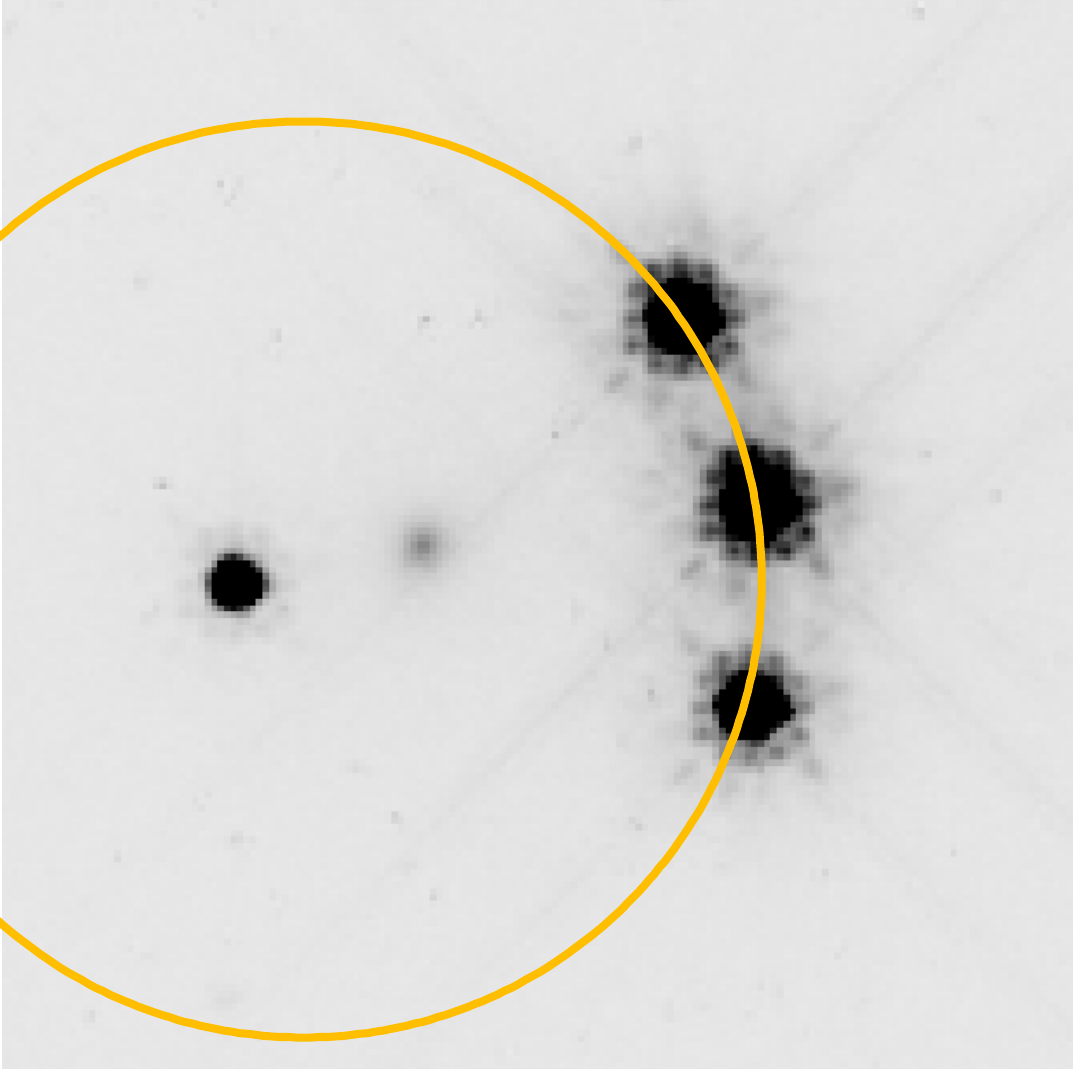}
\caption{An HST image of the lensed quasar system PS J0147+46 with a
  circle drawn through the three closest images, as described in
  Section \ref{subsec:relief}.  The fourth image
  lies well inside the circle, demonstrating the strong
  non-circularity of the system.  In their Figure 1a,
  \citet{Luhtaru_2021}
  show an ellipse passing through all four quasar images.
}
\label{fig:nolocus}
\end{figure}
\begin{figure*}[t!]
\plotone{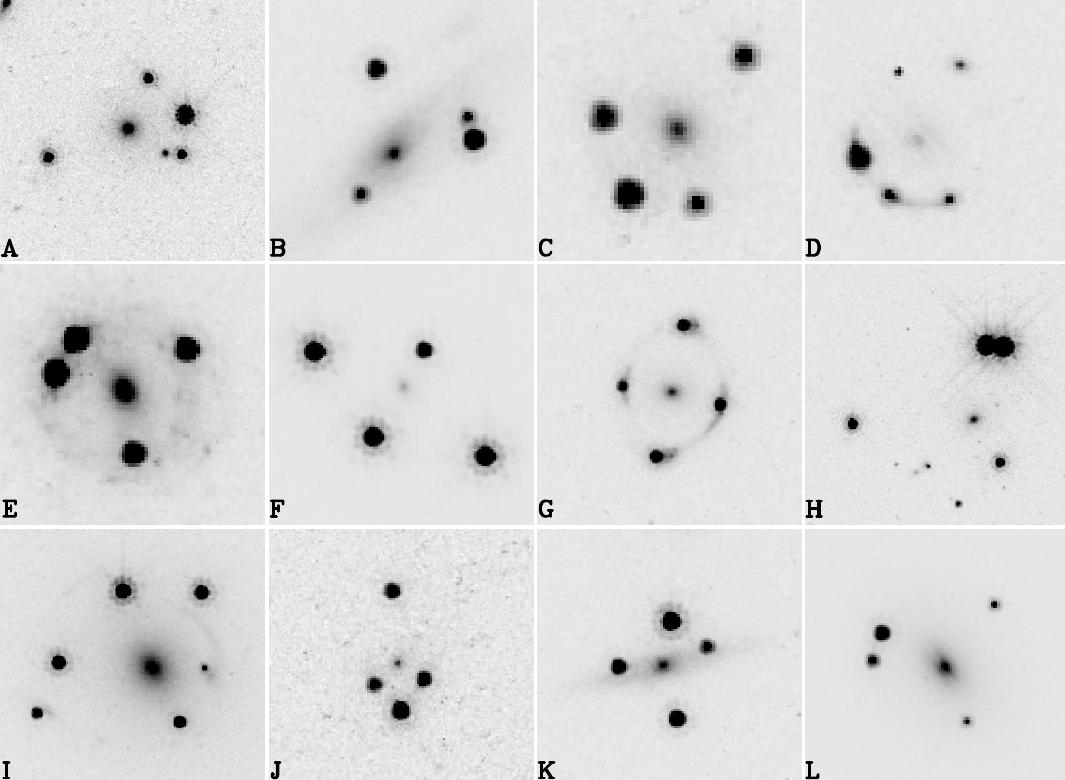}
\caption{HST exposures of a dozen arbitrarily selected quadruply
  lensed quasars.  The reader is invited to classify them
  using the precepts of Section \ref{sec:intro}.  The lensing
  galaxy is always visible at the center of the field.
  The faint background galaxies in panels A and H should be
  disregarded.  Likewise ignore the
  two faint starlike images in panel I and a faint charged
  particle detection in panel D.
  The four {\it brightest} images are those of the quadruply lensed
  quasar.  In panels D, E and G, the galaxy that hosts the
  quasar can be seen forming something of a ring.}
\label{fig:class12}
\end{figure*}
\subsection{Ignore the position of the lensing galaxy}\label{subsec:galpos}

Our classification scheme does not take account of the position of the
lensing galaxy.  It is often not seen in discovery images, both
because it tends to be fainter than the quasar images, and because it
is often crowded by them.

A second reason for ignoring the lensing galaxy is that
it has little effect on the image positions that
drive our classification scheme.  In Appendix \ref{app:expert}
we take the SIEP as our preferred model, but there is
a second model, the singular isothermal sphere with
external shear (henceforth SIS+XS) that predicts
exactly the same image positions.  The two models
predict different positions for the galaxy \citep{Luhtaru_2021},
but our classification scheme is well suited to both.

\section{Precise estimates of ellipticity versus crude guesstimates
  of non-circularity}\label{sec:ellipses}

\subsection{The problem}

The ellipses shown in Figure \ref{fig:thescheme} were derived by
fitting our preferred theoretical model, the SIEP to the positions of
the four images.  They are characterized by a precise ``ellipticity''
obtained from the ratio, $q$, of the short symmetry axis to the long
symmetry axis.

Unfortunately, four image positions do not yield a unique ellipse
without additionally specifying
the direction of the gravitational potential's symmetry axis.
This is illustrated in Figure 2 of
\citet{Wucknitz_2002}, where the ellipses drawn through each
configuration have a range of orientations and a range of
ellipticities.

How then is the classifier expected to determine non-circularity?
Crudely!

Our scheme does not require precise quantification of non-circularity,
but only the binary choice between near-circular and non-circular.
In the author's experience, a quartet of lensed images
for which the SIEP model has axis ratio $0.9 < q < 1 $
looks quite circular, while a quartet for which $ q < 0.85$
looks very non-circular.

Using the letters $C$ and $N$ to designate near-circular
and non-circular, one might use $C$? and $N$? for
systems about which one has doubts. The classifier is
urged to choose between the two, the lack of certainty
notwithstanding.

\subsection{Relief for the ambiguity intolerant}\label{subsec:relief}

Though we stipulate that the classifier choose between circular and
non-circular, we recognize some will want to eliminate all ambiguity.
One can do this using high school geometry.

It takes only three points to draw a circle (with the center where the
bisectors of the chords intersect).  If one draws a circle through
three images of a quadruple configuration, the fourth image will lie 
close to that circle if the system is near-circular.  This works
best if one chooses the three closest images.

Figure \ref{fig:nolocus} shows this construction for the system PS
J0147+46, one of the more non-circular lenses known.  A circle is drawn
through the three closest images.  The fourth image lies well inside
that circle.

\section{Test Drive}\label{sec:testdrive}

We invite readers to decide for themselves whether the present scheme
might be useful.  In Figure \ref{fig:class12} we show an arbitrary
selection of 12 HST images of quadruply lensed quasars.  We believe
that even novices will find it easy to decide upon the kitelikeness of
these systems.

As discussed in the previous section, distinguishing between
near-circular and non-circular systems is more challenging.
Non-circularity is most easily recognized in the doubly symmetric
super-kites, and also fairly obvious in at least some kites, again by
virtue of their symmetry.  It is least obvious in the doubly
asymmetric un-kites.

We urge readers to make a best effort to classify all 12 systems.
Additional material presented in Appendix \ref{app:encircle} may help
in rectifying first timers' mistakes.  

\acknowledgments

The author gratefully acknowledges the work of Raymond Wynne, Richard
Luhtaru and Chirag Falor in laying the foundations for this paper, and
thanks Dominique Sluse for encouragement.  Observations shown in
Figures \ref{fig:thescheme}, \ref{fig:class12} and 4 were taken as
part of programs HST-GO-15320 and HST-GO-15652 carried out with the
NASA/ESA Hubble Space Telescope and obtained from the Space Telescope
Science Institute, which is operated by the Association of
Universities for Research in Astronomy, Inc.  under NASA contract NAS
5-26555.

%

\vspace{5mm}


\appendix

\section{Notes for experts}\label{app:expert}

\begin{figure*}[t]
\plotone{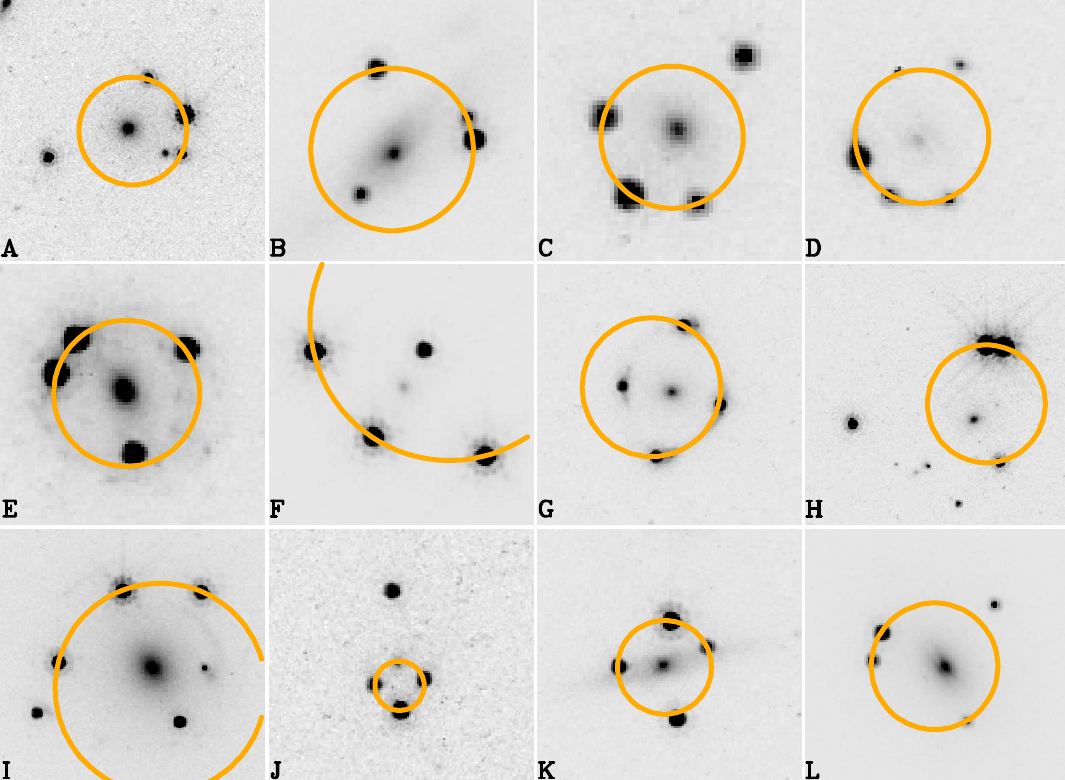}
\caption{
  For each of the systems in our test sample,
  orange circles have been drawn through the three closest images.
  The fourth image in systems D and E, both un-kites, lie very close
  to the circle.  By contrast, in system H, also an un-kite, the fourth
  image lies very far from the circle.  By coincidence,
  a charged particle detection lies on the circle of system D.
}\label{fig:overcircle}
\end{figure*}

The taxonmy presented here was inspired by the work of
\citet{FalorSchechter} who studied the quadruple image configurations
of what they call the ``asymptotically circular lens.''  The
configurations are determined by the position of the center of the
circle within a true astroid, which marks the transition from four
images to two.  The image positions perfectly satisfy the
configuration invariant discovered by \citet{KassiolaKovner}.  They
showed that the invariant is very nearly satisfied by most known
quadruply lensed quasars.

Our three kite classes correspond to having the center of the
circle be a) close to the center of the astroid, b) close
to one of its two axes or c) close to neither axis.

The singular isothermal elliptical potential (SIEP), the singular
isothermal sphere with external shear (SIEP+XS), and the singular
isothermal elliptical potential with parallel shear
(SIEP+XS$_\parallel$) all produce elliptical configurations that are
``scronched'' \citep{ellenberg2021shape} versions of the asymptotically
circular configurations, adding a third dimension and giving
us the distinction between nearly-circular and non-circular
systems.

The simplest scheme for determining the orientation of the lensing
potential on the sky involves solving for Witt's hyperbola
\citep{Witt_1996}, which in turn requires solving a system of 3 linear
equations in 3 unknowns.  As this may be out of reach for many citizen
scientists, we describe in Section \ref{subsec:relief} a simpler,
{\it ad-hoc} construction to quantify the deviation from
non-circularity.

We do not refer to ``crosses'', ``folds'' and ``cusps'' for two
reasons.  First, we have avoided jargon where possible.  Second, these
are limiting cases of our three, more broadly defined kite classes.

We likewise do not refer to the classification scheme developed by
\citep{Saha_2006}.  Their scheme does not explicitly allow for the
elongation of image configurations.  But it does make a distinction
between what they call long-axis quads and short-axis quads,
indicating the axis of the potential on which the source lies.  We
would call these wide and narrow kites, respectively, as source
position is a theoretical concept that we have avoided mentioning.
The distinction between the two is very small except for the
most non-circular systems.  We have therefore chosen not to add this
distinction to our scheme.

Our taxonomy is built on two commonly used and widely understood words
-- the noun kite and the adjective circular.  A system is then
described as a prefix-adjective prefix-noun.  While the word
``rhombus'' might be mathematically more rigorous,
we use super-kite to emphasize
that it is the degree of kitelikeness that distinguishes our three
classes.

\section{Deviations from circularity}\label{app:encircle}

As discussed in Section \ref{sec:ellipses}, ellipses with a range of
orientations and axis ratios can be drawn through four images that are
not perfectly symmetric.  We described a construction that permits a
quantitative estimate of non-circularity.

While our hope is that novices will quickly develop
confidence in their ability to make a crude distinction between
nearly circular and non-circular systems, this construction
may prove useful to first time users who do not yet trust
their own judgement.  We have carried this out for each
the 12 systems in our test sample, with the resulting circles
shown in Figure \ref{fig:overcircle}.

Table 1 reports ellipticities, defined as $\epsilon \equiv 1 - q$,
where $q$ is the axis ratio.  They were obtained from fits of
theoretical SIEP models to the test sample.

\begin{deluxetable}{lrlrlrlr}
\tablecolumns{8}
\tablewidth{0pc}
\tablecaption{SIEP model ellipticities $\epsilon$ for the test sample}
\tablehead{
  \colhead{name}    & \colhead{$\epsilon$} &
  \colhead{name}    & \colhead{$\epsilon$} &
  \colhead{name}    & \colhead{$\epsilon$} &
  \colhead{name}    & \colhead{$\epsilon$} 
}
\startdata
J0659+16 & 0.15 & J1330+18 & 0.12 & J2205$-$37 & 0.15 & J1042+16 & 0.05 \\
J1131$-$44 & 0.06 & J1134$-$21 & 0.50 & J1537$-$30 & 0.27 & J0818$-$26 & 0.40 \\
J1721+88 & 0.21 & J0029$-$38 & 0.52 & J1817+27 & 0.05 & J2100$-$44 & 0.13 \\
\enddata
\end{deluxetable}




\bibliography{schechter}{} \bibliographystyle{aasjournal}



\end{document}